\documentstyle[12pt,epsf]{article}\textheight
230mm\textwidth 165mm 
\newcommand{\bi}{\bibitem}
\newcommand{\be}{\begin{eqnarray}}
\newcommand{\ee}{\end{eqnarray}}
\newcommand{\nn}{\nonumber}
\catcode`\@=11
\def\lsim{\mathrel{\mathpalette\@versim<}}
\def\gsim{\mathrel{\mathpalette\@versim>}}
\def\@versim#1#2{\vcenter{\offinterlineskip
\ialign{$\m@th#1\hfil##\hfil$\crcr#2\crcr\sim\crcr } }}
\catcode`\@=12

\begin{document}
\pagestyle{empty}

\begin{center}
{\Huge\bf  Applications of the Reduction of Couplings$^{*}$}
 \end{center}

\vspace{0.5cm}
\begin{center}
{\large Dedicated to 
Professor Wolfhart Zimmermann \\ on the occasion 
of his 70th birthday}
\end{center} 

\vspace{2cm}
\begin{center}
{\sc Jisuke Kubo}
\end{center}
\begin{center}
{\em 
Institute for Theoretical Physics, 
Kanazawa  University\\
Kanazawa 920-1192, Japan}
\end{center}

\vspace{1cm}
\begin{center}
{\bf Abstract}
\end{center}

\noindent
Applications of the principle of reduction of couplings
to the standard model and supersymmetric grand
unified theories are reviewed.
Phenomenological applications of renormalization group invariant
sum rules for soft supersymmetry-breaking parameters are
also reviewed.

\vspace*{3cm}
\footnoterule
\vspace*{2mm}
\noindent
$^{*}$ To appear in the Proceedings of 
the Ringberg Symposium on {\em Quantum Field Theory},
Tegernsee, Germany, June 1998.

\newpage
\pagestyle{plain}

\section{ Application to the Standard Model}

High energy physicists have been using renormalizability
as the predictive tool, and also to decide whether or not
a quantity is calculable.
As we have learned in the previous talk by Professor Oehme,
it is possible, using the method
of reduction of couplings \cite{zim1,zim2,oehme1}, to renormalize a theory 
with fewer number of counter terms then
 usually counted, implying that
the traditional notion of renormalizability should be 
generalized in a certain sense \footnote{Earlier
references related to the idea of reduction
of couplings are given in \cite{chang1}. 
Professor Shirkov who is present
here was one of the authors 
who considered such theoretical possibilities. For reviews, see
\cite{review}.}.
Consequently, the notion of the predictability and
the calculability  \cite{georgi1} may also be generalized
with the help of reduction of couplings.
Of course, whether the generalizations of these notions
have anything to do with  nature is another question.
The question can be answered if one
applies the idea of reduction of couplings to realistic
models, make predictions that are specific
for reduction of couplings, and then  wait
till experimentalists find positive results
\footnote{Here I would like to
restrict myself to phenomenological applications of 
reduction of couplings. 
See \cite{kmz3,ref-after} for the other applications.
Professor Oehme reminded me that
in his seminar talk given at
the Max-Planck-Institute early 1984, professor Peccei suggested 
phenomenological applications of this idea .}.

In 1984
Professor Zimmermann,  Klaus Sibold and myself  \cite{kubo1}
began to apply the idea of reduction of couplings to
the  $SU(3)_{\rm C}\times SU(2)_{\rm L} 
\times U(1)_{\rm Y}$ gauge model
for the strong and electroweak interactions.
As it is known, this theory has a lot of free parameters,
and at first sight it seemed there exists no guiding principle how
to reduce the couplings in this theory.
There were two main problems associated with this program.
The one was that it is not possible to 
assume a common asymptotic behavior for  all couplings,
and the other one is how to increase the predictive 
power of the model without
 running into the contradiction with the experimental 
knowledge (of that time) such as the masses of the known fermions.
Professor Zimmermann suggested to use
asymptotic freedom  as a guiding principle,
and assumed that QCD is most fundamental among the interactions
of the standard model (SM).
Since pure QCD is asymptotically free, we  tried to 
switch on  as many SM interactions as possible 
while keeping asymptotic freedom
and added them  to QCD. The result was almost unique:
There exist two possibilities (or two 
asymptotically free (AF) surfaces in the space of couplings).
It turned out that on the first surface, 
the $SU(2)_{\rm L}$ gauge coupling  $\alpha_2$
is bigger than the QCD coupling  $\alpha_3$,
and on the second surface, $\alpha_2$
has to identically vanish.
We decided to choose the second possibility, because
we found out that
it is possible to include the $SU(2)_{\rm L}$
gauge coupling $\alpha_2$
as a certain kind of ``perturbation'' into the AF system.
Since the perturbating couplings should be regarded as free parameters,
the reduction of couplings in this case can be achieved only
partially (``partial reduction'').
For the first case, it was not possible.

Thus, the largest AF system 
which is phenomenologically acceptable (at that time) contains
$\alpha_3$, 
the quark Yukawa couplings $\alpha_{i}~~(i=d,s,b,u,c,t)$
and the Higgs self coupling $\alpha_{\lambda}$.
However, because of the hierarchy of the
Yukawa couplings,
we could not expect that all couplings can be expressed
in terms of a power series of $\alpha_3$
without running into the contradiction with
that hierarchy.
So we decided to apply to the reduction of couplings only
to the system with 
$\alpha_3~,~\alpha_{t}$ and $\alpha_{\lambda}$, 
and to regard the other couplings as perturbations like $\alpha_2$.
 \begin{figure}
           \epsfxsize= 11 cm   
            \centerline{\epsffile{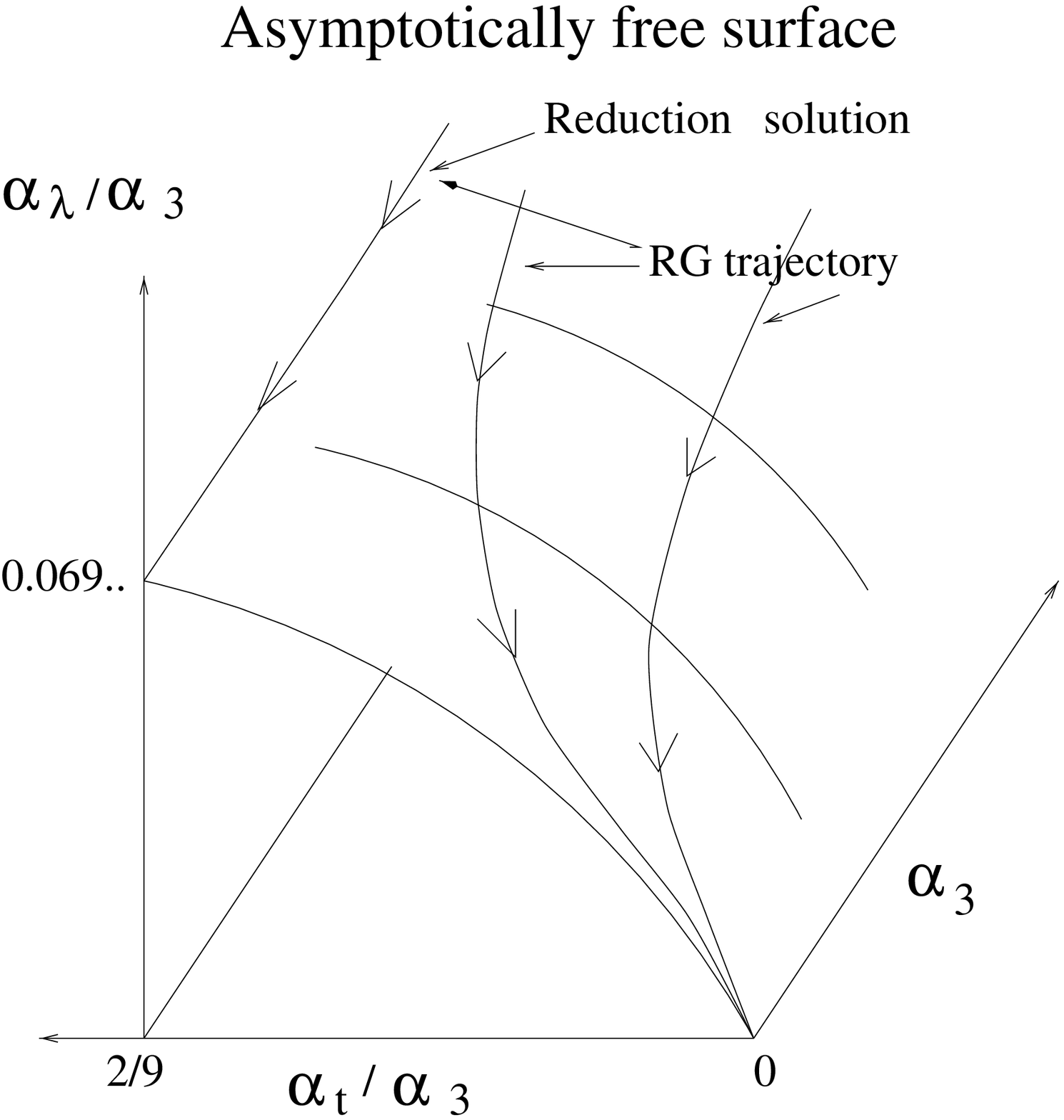}}
         \caption{Reduction of $\alpha_t$ and $\alpha_{\lambda}$
  in favor of $\alpha_3$.}
         \label{fig: 1}
         \end{figure}
Fig. 1 shows the AF surface in the space of
$\alpha_3$, $\alpha_{t/\alpha_3}$ and  $\alpha_{\lambda}/\alpha_3$.
The reduction of the top Yukawa and  Higgs
couplings in favor of the QCD coupling corresponds to the 
border line on the surface, i.e.,  the line defined by
\be
\alpha_{t}/\alpha_3 &=& \frac{2}{9}~,~
\alpha_{\lambda}/\alpha_3=\frac{\sqrt{689}-25}{18}\simeq 0.0694~ 
\ee
in the one-loop approximation.
This border line was
already known as the Pendleton-Ross infrared (IR) 
fixed point (line) \cite{pendleton1}.
Note that the existence of the AF surface (shown in Fig. 1) at least for 
$\alpha_3$ closed to the origin is mathematically
ensured (see also \cite{zim3}), while the line
for large $\alpha_3$, Pendleton-Ross 
infrared IR line, can be an one-loop artifact which was pointed out
by Professor Zimmermann. He showed explicitly in the two-loop
approximation that this is  indeed  the case \cite{zim4}.

An asymptotically free renormalization group (RG) trajectory lies 
exactly on the surface. 
 \begin{figure}
            \epsfxsize= 11 cm   
            \centerline{\epsffile{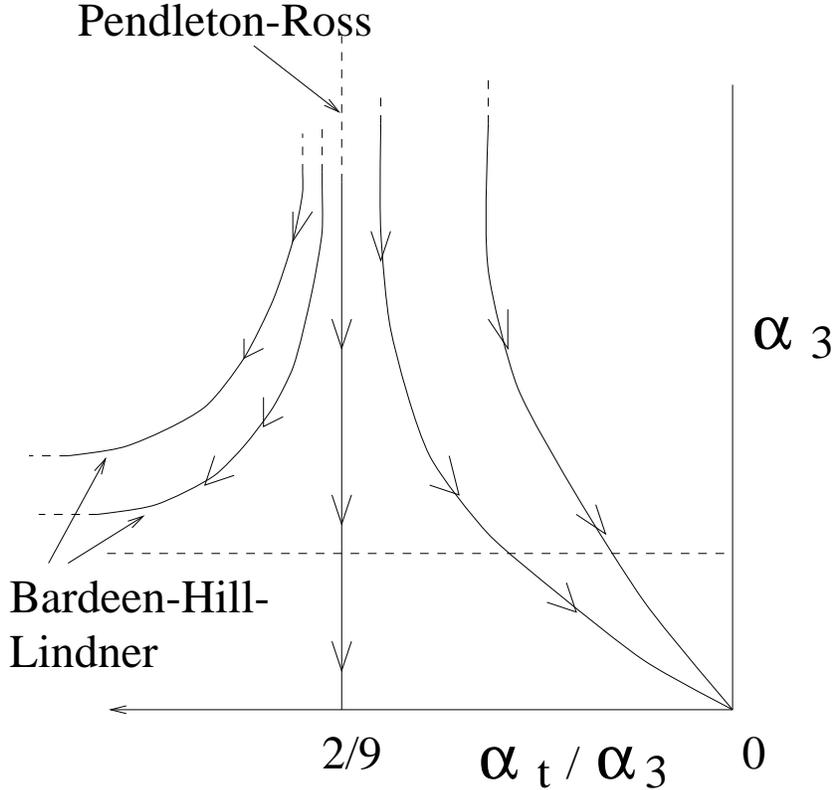}}
         \caption{Asymptotically free surface in the 
    $\alpha_3-\alpha_{t}/\alpha_3$ space.}
         \label{fig: 2}
         \end{figure}
Fig. 2 shows  trajectories projected on the 
$\alpha_3-\alpha_{t}/\alpha_3$ 
plane.
It may be worthwhile to mention that the branches above the
Pendleton-Ross IR line (the lines left to it in Fig. 2) 
are used by Professor
Bardeen and his collaborators \cite{bardeen1} to interpret the Higgs
particle as a bound state of the top and anti-top quarks.
From Fig. 2 one can see that the higher the energy scale
where  the top Yukawa coupling 
diverges (the horizontal dotted line in Fig. 2 will be
lowered), the similar
is the prediction of the top mass in two methods. However, I would like to
emphasize that how to include the corrections to this
lowest order system (especially those due to
 the non-vanishing $SU(2)_{\rm L}$
and $U(1)_{\rm Y}$ gauge couplings) depends on the ideas behind,
so that the actual predictions are different.
We  included these corrections within the one-loop approximation
and  calculated $\alpha_{t}/\alpha_{3}$ and $\alpha_{h}/\alpha_{3}$
in terms of $\alpha_{3}$ and the perturbating free couplings. Then we
used the formulae
\be
M_{t}^{2}/M_{Z}^{2}\ =\ 2\cos^{2}\theta_{\rm W}
\alpha_{t}/\alpha_{2} ~,~M_{h}^{2}/M_{Z}^{2}\ =\ 2\cos^{2}\theta_{\rm W}
\alpha_{h}/\alpha_{2}~,
\ee
to calculate the top quark and Higgs masses, 
$M_{t}$ and $M_{h}$, from the known values of the parameters such
as the $Z$ boson mass $M_{Z}$ and the Weinberg
mixing angle  $\theta_{\rm W}$. We obtained \cite{kubo1}
 \be
M_{t} & \simeq & 81 ~~\mbox{GeV}\ , 
M_{h}  \simeq  61 ~~\mbox{GeV}~.
\label{top85}
\ee
Later I included higher order corrections such
as two-loop corrections and found that the earlier predictions
(\ref{top85}) become
$M_{t} = 98.6\pm9.2$ GeV and $M_{h} = 64.5\pm1.5$ GeV,
which should be compared  with the present knowledge \cite{pdg}
\be
M_{t} &=& 173.8\pm 5.2 ~~\mbox{GeV}\ ,
M_{h} \gsim 77.5~~\mbox{GeV}\ .
\ee

The failure of our prediction was disappointing in fact.
However, this failure relieved Professor Zimmermann
from a self-contradicting feeling.
As we know he likes low-energy supersymmetry
and also good wines.
If our prediction would have been confirmed by an experiment,
it would be very unlikely that low-energy supersymmetry
is realized in nature, which would imply that
he would loose {\em again} a lot of bottles of wines.
So, the decision of nature was welcome at the same time.
Fig. 3 summarizes.
 \begin{figure}
            \epsfxsize= 11 cm   
            \centerline{\epsffile{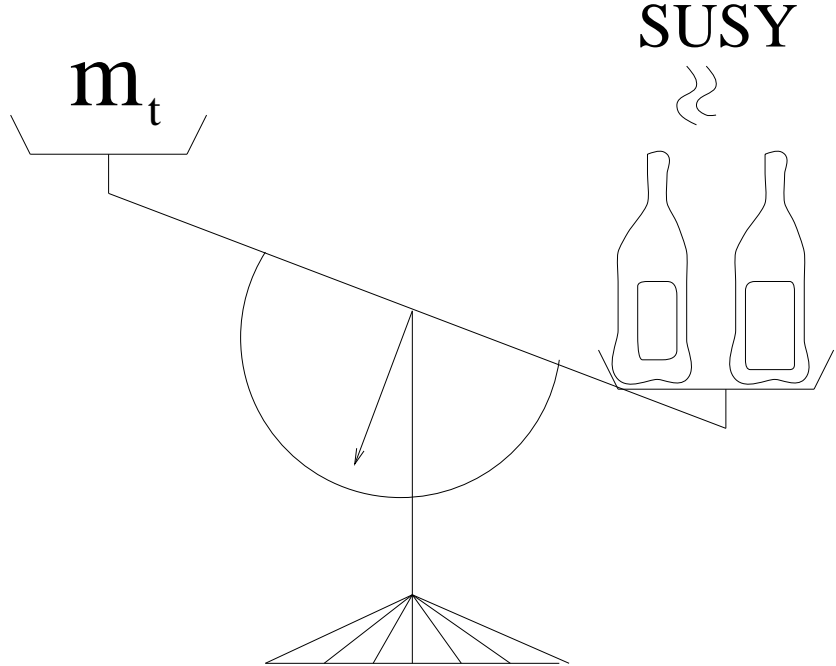}}
         \caption{Entt\" auschung und Hoffnung.}
         \label{fig: 3}
         \end{figure}

\section{Why is Supersymmetry as Ideal Place for Application?}
\subsection{Naturalness and supersymmetry}
Let me now come to the application of reduction of parameters
to supersymmetric theories.
I do not know why Professor Zimmermann likes low energy
supersymmetry. But let me assume that he likes
the usual argument for  low energy supersymmetry, which is based on 
the naturalness notion of 't Hooft \cite{thooft1}.
I would like to spend few minutes for that.
(Let me allow to do so, although
for the superexperts in the audience it might be superboring.)
't Hooft  \cite{thooft1} said that there exist a natural scale in a given
theory, and that the natural energy scale of spontaneously broken gauge
theories which contain the SM is usually less than few TeV.
The argument is the following. 
Suppose the scale at which the SM goes over to a more fundamental
theory is $\Lambda$. That is, there are in the 
fundamental theory particles
with masses of this order.
Now consider the propagator  $\Delta (p^2)$
of a boson field with the physical  mass $m_{\rm B}$ 
much smaller than  $\Lambda$,
and suppose that it
 is
normalized at $\Lambda$  so that the propagator assumes a
simple form at $p^2=-\Lambda^2$:
\be
\lim_{p^2\to -\Lambda^2}\Delta (p^2) 
&\to & \frac{i Z(\Lambda^2)}{p^2-m^{2}_{\rm B}(\Lambda^2)}~,
\ee
where $Z$ is the normalization constant for  the wave function.
The physical mass squared $m^{2}_{\rm B}$ can be expressed as
\be
m^{2}_{\rm B} &=& m^{2}_{\rm B}(\Lambda^2)+\delta m^{2}_{\rm B}~.
\ee
Then we ask ourselves how accurate we have to tune the value of
$m_{\rm B}^{2} (\Lambda^2)$ to obtain a desired accuracy
in the physical mass squared $m^{2}_{\rm B}$.
This depends on $\delta m^{2}_{\rm B}$, of course.
't Hooft said that for a theory to be natural
the ratio $ m^{2}_{\rm B}(\Lambda^2)/ m^{2}_{\rm B}$
 should be of $O(1)$,
which implies that $|\delta m_{\rm B}^{2}| < m_{\rm B}^{2}$.
If quadratic divergences are involved in 
the theory, the correction $\delta m_{\rm B}^{2}$
will be proportional  not only to the masses of the light fields,
but also to the masses of the heavy fields, and so $\delta m_{\rm B}^{2}$
can be of the order $(\alpha/4\pi) \Lambda^2$, where 
$\alpha$ is some generic coupling.
Since the Higgs mass should not exceed
few hundred GeV in the SM, 
the natural scale of the fundamental
theory, which contains the standard model Higgs  and
also involves quadratic divergences, is at best few TeV.
So according 't Hooft, ordinary Grand Unified Theories (GUTs),
 for instance,
are unnatural \cite{thooft1}.

Supersymmetry, thanks to its very renormalization 
property known as  non-renormalization
theorem \cite{nonre1,nonre2}, can save the situation.
The cancellation of the quadratic divergences, which
was first observed by Professors Wess and Zumino \cite{nonre1},
is exact if the masses of the bosonic and fermionic superpartners
are the same.
However, supersymmetry is unfortunately broken in nature,
so that the cancellation is not exact.
The mass squared difference, $m_{\rm B}^{2}-m_{\rm F}^{2}$,
 characterizes the energy scale of 
supersymmetry breaking.
To make compatible supersymmetry breaking
with the naturalness notion of 't Hooft, we must impose
the constraint on the supersymmetry-breaking scale $M_{\rm SUSY}$.
A simple calculation yields that $M_{\rm SUSY}$ should
be less than few TeV.

\subsection{Soft supersymmetry-breaking parameters}
Since the pioneering works by Professor Iliopolos 
(who could not participate in this meeting) 
with
P. Fayet \cite{iliopoulos1} and the others in late 70's, a
lot of attempts to understand supersymmetry-breaking 
mechanism have been
done. However, unfortunately,
 we still do not know
how supersymmetry is really broken in nature.  It, therefore, may be
reasonable
at this moment to pick up the common feature 
of  supersymmetry breaking which effect the SM.
The so-called minimal supersymmetric standard model (MSSM) is ``defined''
along this line of thought.
The MSSM contains the ordinary gauge bosons and fermions 
together with their superpartners, and two supermultiplets
for the Higgs sector. (With one supermultiplet
in the Higgs sector, it is not possible to give 
masses to all the fermions of the MSSM.)

It is expected that the common effect
of supersymmetry breaking is to add the so-called 
soft supersymmetry-breaking terms (SSB) to the symmetry theory.
The SSB terms
are defined as those which do not change the infinity structure
of the parameters of the symmetric theory.
So they are additional
terms in the Lagrangian that do not change
the RG functions such as the $\beta$- and $\gamma$-functions
of the symmetric theory.
(More precisely, there exists a renormalization scheme
in which the RG functions are not altered
by the SSB terms.)
There exist four types of such terms \cite{soft}.
\be
1. & &\mbox{Soft scalar mass terms}:~~~
(m^2)_i^j \phi_j \phi^{* i}~,\nn\\
2. & & B-\mbox{terms}:~~~
B^{ij} \phi_j \phi_{i} +~\mbox{H.C}~,\\
3. & & \mbox{Gaugino mass terms}:~~~
M \lambda\lambda +~\mbox{H.C}~,\nn\\
4. & & \mbox{Trilinear scalar  couplings}:~~~
h^{ijk}\phi_i \phi_j\phi_k+~\mbox{H.C}~,\nn
\ee
where $\phi_j$ and $\lambda$ denote the scalar component
in a chiral supermultiplet and the gaugino 
(the fermionic component) in a gauge
supermultiplet, respectively.

If one insists only renormalizability for the
MSSM, the number of the SSB parameters 
amounts to  about $100$, which is 
about  five times of that of the SM.
The commonly made assumption to reduce
this number is the assumption of 
universality of the SSB terms,
which is often justified by saying that
supersymmetry breaking occurs in a flavor blind sector \cite{nilles1}.
That is, it is assumed that
the soft scalar masses and the trilinear scalar couplings 
are universal or flavor blind at the
scale where supersymmetry breaking takes place.
The so-called constrained MSSM contains thus
only four independent massive parameters.
But we could easily  imagine that nature might not be so universal
as one wants.
In fact it possible to construct a lot of models 
with non-universal SSB terms \cite{dixon1} (even in models in which 
supersymmetry-breaking occurs in the so-called hidden sector
which does not interact directly with the observable sector), and 
once we deviate from the universality, there will be
chaotic varieties.

The application of reduction of couplings
in the SSB sector is based on 
the assumption that
the SSB terms organize themselves into a most economic
structure that is consistent with renormalizability.
I will come to discuss this later.
I have spent a lot of time for 
low energy supersymmetry, because I wanted to  argue that
supersymmetric theories offer
an ideal place where the reduction method, especially for massive
parameters, can be applied and tested experimentally.
It is worthwhile to mention that
the current research program of Professor Zimmermann
is the reduction of massive parameters \cite{zim5}.

\section{Supersymmetric Gauge-Yukawa Unification}
Before I come to discuss the SSB sector, I would like to stay
in the sector of the dimensionless couplings in
 realistic
supersymmetric GUTs and tell about certain phenomenological successes of 
reduction of parameters in these theories.
I would like to emphasize  that in contrast to the SM, 
supersymmetric GUTs can be
asymptotically free or even finite.

\subsection{Unification of the gauge and
Yukawa couplings based on the principle of reduction
of couplings}
Few year ago, Professor Zimmermann and I were trying
to apply the reduction method in the dimensionless sector
of the MSSM, but we had no success.
The main reason was that the power series solution to the
reduction equation seemed to diverge.
So we stopped to continue.
About the same time, George Zoupanos (who unfortunately
could not come here today) visited the Max-Planck-Institute,
and told me that he obtains a top quark mass of
about 180 GeV in a finite $SU(5)$ GUT \cite{finite1}.
Although the top quark was not found at that time 
(it was end of 1993, 
so just before we heard the rumor from Fermilab),
180 GeV for the top quark mass was a reasonable value.
Finite theories have attracted
many theorists. 
By a finite theory
we mean a theory with the vanishing
$\beta$-functions and anomalous dimensions.
As we know, the $N=4$ 
supersymmetric Yang-Mills theory is 
a well-known example \cite{n=4}.
And there were many attempts to construct $N=1$
supersymmetric finite theories  
\cite{finite1,parkes1,kkmz1}. 
Klaus Sibold and his collaborators \cite{lucchesi1}
gave an elegant existence proof
of finite $N=1$ supersymmetric theories,
where I would like to recall that
their proof is strongly based on 
the Adler-Bardeen non-renormalization theorem
of chiral anomaly \cite{adler1} (about which Professor Bardeen 
talked  yesterday) \footnote{It is currently studied 
how to extend their theorem; for instance
a non-perturbative extension 
has also been proposed in \cite{strassler1}. }.
The reduction of Yukawa couplings in favor
of the gauge coupling
 is one of the necessary condition
for a theory to be finite in perturbation theory.
So in a finite theory,
Gauge-Yukawa unification 
is achieved.
Since Gauge-Yukawa unification results
from the reduction of Yukawa couplings in favor of the gauge
coupling,
it can be achieved
not only in finite theories but also
in non-finite theories, as Myriam Mondrag\' on, 
George Zoupanos and myself explicitly showed \cite{kmz1}.
Relations among the gauge and Yukawa couplings,
which are missing in ordinary GUTs,
could be a consequence of a further unification provided by a more 
fundamental theory. And so Gauge-Yukawa unification is a natural
extension to the ordinary GUT idea.
This  idea of  unification relies  
on a symmetry principle as well as on the principle of
reduction of couplings. The latter principle requires the
existence of RG invariant relations among couplings, which 
do not necessarily result from a symmetry,
but nevertheless preserve perturbative renormalizability
or even finiteness  as I  mentioned.

\subsection{The double-role of $\tan\beta$}
Before I come to discuss Gauge-Yukawa unification
more in details, I would like to talk about an important
parameter, $\tan\beta$, in the MSSM. It is a very popular 
parameter among SUSY
 physicists, but let me allow to spend few minutes for this parameter,
because it plays also an important role for Gauge-Yukawa unification.
As I mentioned the MSSM contains two Higgs
supermultiplets.
The most general form of the Higgs potential
which is consistent with renormalizability
and with the softness of the SSB parameters can be written as
\be
V &=&
(m_{H_d}^{2}+|\mu_H|^2)\,\hat{H}_{d}^{\dag}\hat{H}_{d}+
(m_{H_u}^{2}+|\mu_H|^2)\,\hat{H}_{u}^{\dag}\hat{H}_{u}+
(\,B\hat{H}_{d}\hat{H}_{u}+~\mbox{H.C.}\,)\nn\\
& &+\frac{\pi}{2}( 3 \alpha_{1}^{2}/5
+\alpha_{2}^{2})(\, \hat{H}_{d}^{\dag}\hat{H}_{d}-
\hat{H}_{u}^{\dag}\hat{H}_{u}  \,)^2 ~,
\label{potential1}
\ee
where $\mu_H$ is the only 
massive parameter in the supersymmetric limit, while
$m_{H_u}^{2}$ , $m_{H_d}^{2}$ and $B$ 
are the SSB parameters in this sector.
( $m_{H_u}^{2}$ , 
$m_{H_d}^{2}$ are real while $\mu_H$ and $B$ may be complex
parameters.)
Here $\hat{H}_{u,d}$ denote the scalar components of the two
Higgs supermultiplets.
There are four independent massive parameters in this sector
as we can see in (\ref{potential1}).
These  parameters should give 
the only one independent mass parameter of the SM, 
for instance the mass of Z.
Now instead of regarding these parameters as independent one can
regard also the ratio of the vacuum expectation values \cite{inoue1}
\be
\tan\beta &\equiv &\frac{<\hat{H}_{u}>}{<\hat{H}_{d}>}
\label{tanb}
\ee
as independent. ($\tan\beta$ can be assumed to be  real.)
Usually one regards $|\mu_H|$ and
$B$ as dependent \footnote{If $\tan\beta$ is real
as we assume here, $B$
can become complex starting in one-loop order \cite{pilaftsis}.}.
So the Higgs sector in the tree approximation is characterized by the
parameters \be
\tan\beta~,~m_{H_1}^{2}~,~m_{H_2}^{2}~.
\ee
The crucial point  for Gauge-Yukawa unification
is that $\tan\beta$ plays a double-role.
On one hand, it is a parameter in the Higgs potential
 as we have seen above, 
and on the other hand it it is a mixing parameter
to define the standard model Higgs field
out of the two Higgs fields of the MSSM.
That is, $\tan\beta$ appears also in the dimensionless
sector, and in fact it can be fixed through
Gauge-Yukawa unification with the knowledge of the tau mass $M_{\tau}$, 
as I would like to
explain it more in detail below.

\subsection{How to predict $M_{t}$ from 
Gauge-Yukawa Unification}
The consequence of a Gauge-Yukawa unification in a GUT is that 
the gauge and Yukawa couplings  are related above the
GUT scale $M_{\rm GUT}$.
In the following discussions we consider only the Gauge-Yukawa
unification in the third generation 
sector  \footnote{A naive extension to include other
generations into this scheme fails phenomenologically.}:
\be
g_i& = &\kappa_i \,g~\sum_{n=1}(~1+\kappa_{i}^{(n)}
 g^{2n}~)~~
~(i=1,2,3,t,b,\tau)~,
\label{gyu2}
\ee
where $g$ denotes the unified gauge coupling,
$g_i$ denote the gauge and Yukawa
couplings of the MSSM.
Note that the constants $\kappa_i$'s can be explicitly
calculated from the principle of reduction of
couplings.
Once $\tan\beta$ and the Yukawa couplings are
known, the  fermion
masses  can be calculated as one can easily see from the tree level mass formulae
\be
M_t &=& \sqrt{2}\frac{M_{\rm Z}}{g_2}\sin\beta\cos\theta_{\rm W}~g_t ~,
~ M_{b,\tau}=\sqrt{2}\frac{M_{\rm Z}}{g_2}\cos\beta
\cos\theta_{\rm W}~g_{b,\tau}~,
\ee
where $M_t$, $M_b$ and $M_{\tau}$ are the masses of the top and 
bottom quarks and tau, respectively.
Assume  that we use the tau mass $M_{\tau}$ 
as input and also that 
below $M_{\rm SUSY}~(>M_{t})$ the effective theory
of the GUT is the SM. At $M_{\rm SUSY}$ the couplings
of the SM and MSSM have to satisfy
 the matching conditions
\footnote{There are MSSM threshold corrections
to the matching conditions \cite{hall1,wright1}, 
which are ignored here.}
\be
\alpha_{t}^{\rm SM} 
&=&\alpha_{t}\,\sin^2 \beta~,~
\alpha_{b}^{\rm SM}
~ =~ \alpha_{b}\,\cos^2 \beta~,
~\alpha_{\tau}^{\rm SM}
~=~\alpha_{\tau}\,\cos^2 \beta~,\nn\\
\alpha_{\lambda}&=&
\frac{1}{4}(\frac{3}{5}\alpha_{1}
+\alpha_2)\,\cos^2 2\beta~~~(\alpha_i=\frac{g_i^2}{4\pi})~,
\label{matching}
\ee
where 
 $\alpha_{i}^{\rm SM}~(i=t,b,\tau)$ are
the SM Yukawa couplings and $\alpha_{\lambda}$ is the Higgs coupling.
It is now easy to see that there is no longer  freedom 
for $\tan\beta$ because with a given set of the input
parameters, especially 
$M_{\tau} =1.777$ GeV and $M_Z=91.187$ GeV, 
the matching conditions (\ref{matching}) at  $M_{\rm SUSY}$
and the Gauge-Yukawa unification
boundary condition (\ref{gyu2}) at $M_{\rm GUT}$ can be 
simultaneously satisfied only if we have a
specific value of $\tan\beta$.
In this way
Gauge-Yukawa unification  enables us to predict
the top and bottom masses in 
supersymmetric GUTs.

Table 1 shows the predictions
in the case of a finite $SU(5)$ GUT \cite{kkmz1}, in which
the one-loop reduction solution is given by
\be
g_t^2& = &\frac{4}{5} \,g^2~,~g_b^2=g_{\tau}^{2}
 = \frac{3}{5} \,g^2~.
\label{gyu3}
\ee
\begin{table}
\caption{The predictions 
for different $M_{\rm SUSY}$ for the finite
$SU(5)$ model.}
\begin{center}
\begin{tabular}{|c|c|c|c|c|c|}
\hline
$M$ [GeV]   &$\alpha_{3}(M_Z)$ &
$\tan \beta$  &  $M_{\rm GUT}$ [GeV] 
 & $M_{b}$ [GeV]& $M_{t}$ [GeV]
\\ \hline
$800$ & $0.118 $  &$48.2 $  & $1.3\times 10^{16}$
 & $5.4$  & 173\\ \hline
$10^3$ & $0.117 $  &$48.1 $  & $1.2\times 10^{16}$
 & $5.4$  & 173 \\ \hline
$1.2 \times 10^3$ & $0.117 $  &$48.1 $  & $1.1\times 10^{16}$
 & $5.4$  & 173 \\ \hline
\end{tabular}
\end{center}
\end{table}
The experimental value of $M_t$, $M_b$ and 
$\alpha_{3}(M_Z)$ 
are \cite{pdg}
\be
\alpha_{3}(M_Z) &=&
0.119\pm0.002~,~M_t=173.8\pm 5.2 ~~\mbox{GeV}~,~
M_b=5.2 \pm 0.2~~\mbox{GeV}~.
\ee
We see that the predictions of the model reasonably agree 
with the experimental 
values\footnote{The 
correction to $M_b$ coming from the MSSM superpartners 
 can be as large as 50\%
for very large values  of $\tan\beta$ \cite{hall1,wright1}.
In Table 1 we have not not included these corrections
because they depend on the SSB parameters.
The GUT threshold correction are ignored too.}.
This means among other things that the top-bottom hierarchy 
could be
explained to a certain extent in this Gauge-Yukawa unified model,
which should be compared with how
the hierarchy of the gauge couplings of the SM
can be explained if one assumes  the existence of a unifying
gauge symmetry at $M_{\rm GUT}$ \cite{gqw}. 
 More details on the
different gauge-Yukawa unified models 
and their  predictions can be found in \cite{kmz3,shoda,kmoz1}.

\section{Reduction of Massive Parameters:
Application to the Soft Supersymmetry-Breaking Sector}
To formulate  reduction of massive parameters,
one first has to formulate reduction of dimensionless
parameters in a massive theory,
which was initiated by Klaus Sibold and 
his collaborator Piguet \cite{piguet1},
about ten years ago. To keep 
the generality of
the formulation in the massive case  is much more involved than in the
massless case, because the RG functions now can depend on the ratios
of mass parameters in a complicated way.
In the massless case they are just power series in coupling
constants (at least in perturbation theory).
For phenomenological and also practical applications
of the reduction method, it is therefore most convenient
to work in a mass independent renormalization scheme,
such as the
dimensional renormalization scheme.
There  exists a transformation
of one scheme to another one, which was in fact proven first
by Dieter Maison in the $\phi^4$ theory
as far as I am informed, but not published.
As I mentioned,
the current research program of Professor Zimmermann
is to include into the reduction program the massive parameters.
He has already succeeded to carry out the program
in the most general case and is able to 
show the renormalization 
scheme independence of the reduction method \cite{zim5}.
Consequently, there exist a transformation of a set of 
the reduction solutions in a mass-dependent renormalization scheme
into a set of the reduction solutions in a mass-independent
 renormalization scheme,
which generalizes the unpublished result
of Dieter Maison. 
Thus, the naive treatment on the massive parameters (which
was performed in  phenomenological analyses \cite{jack1,kmz2})
can now be justified by his theorem \footnote{It is
assumed in the theorem that the $\beta$-functions in a mass-dependent
renormalization scheme have a sufficiently smooth behavior in
the massless limit \cite{zim5}.}.

\subsection{Application to the minimal model}

Now I would like to come to the SSB sector of a supersymmetric GUT.
Recall that
the Higgs potential (\ref{potential1}) 
(in the tree approximation) is completely characterized by 
the soft scalar masses $m_{H_u}^2$,  $m_{H_d}^2$ and $\tan\beta$,
where 
$\tan\beta$ is fixed through Gauge-Yukawa unification 
as we have seen before. We \cite{kmz2} applied the 
the reduction methode of massive parameters to the SSB sector
of the minimal supersymmetric $SU(5)$ GUT 
with Gauge-Yukawa unification in the third 
generation ($g_t^2=(2533/2605)  g^2~,
~g_b^2=g_{\tau}^{2} = (1491/2605)  g^2$) \cite{kmz1}, and 
obtained the reduction solution
\be
h_t &=&-g_t\,M~,~h_b =-g_b\,M~,
\label{trilinear1}\\
m_{H_u}^{2} &=&-\frac{569}{521} M^{2}~,~
m_{H_d}^{2} =-\frac{460}{521} M^{2}~,\nn\\
m_{ b_R}^{2} & = &
m_{ \tau_L}^{2}=m_{ \nu_{\tau}}^{2}=
\frac{436}{521} M^{2}~,\nn\\
m_{ d_R}^{2} & = &
m_{ e_L}^{2}= m_{ \nu_{e}}^{2}=
m_{ s_R}^{2}  = 
m_{ \mu_L}^{2}= m_{ \nu_{\mu}}^{2}=\frac{8}{5} M^{2}~,
\label{softmass1}\\
m_{ t_L}^{2} & = &m_{ b_L}^{2}=
m_{ t_R}^{2}=m_{ \tau_R}^{2} =\frac{545}{521} M^{2}~,\nn\\
m_{ u_L}^{2} & = &m_{ d_L}^{2}=
m_{ u_R}^{2}=m_{ e_R}^{2} =
m_{ c_L}^{2}  = m_{ s_L}^{2}=
m_{ c_R}^{2}=m_{ \mu_R}^{2}= \frac{12}{5} M^{2}~\nn
\ee
in the one-loop approximation,
where $h_i$'s are the trilinear scalar couplings,
$m_{i}$'s are the soft scalar masses, and $M$ is 
the unified gaugino mass. We found moreover that we can consistently
regard $\mu_H$ and $B$ as free parameters.
As we can see from (\ref{trilinear1}) and
(\ref{softmass1})  the unified gaugino mass parameter $M$ plays 
a similar role as the gravitino mass $m_{2/3}$ in
supergravity coupled to a GUT and characterizes
the scale of the supersymmetry-breaking \footnote{See for instance
\cite{nilles1}.}.
Note that the reduction solution for the soft scalar 
masses (\ref{softmass1}) is not of the universal form while
those for the trilinear couplings  (\ref{trilinear1}) are universal 
in the one-loop approximation.

Regarding the reduction solutions
 (\ref{trilinear1}) and (\ref{softmass1}) as boundary conditions
at $M_{\rm GUT}$ in the minimal supersymmetric GUT
with Gauge-Yukawa unification in the third 
generation, we
can compute the spectrum of the superpartners of the MSSM,
which is shown in Table 2, where we have used the unified gaugino mass
$M =0.5$ TeV.
\begin{table}
\caption{The prediction of the superpartner spectrum 
for  $M=0.5$ TeV in the minimal gauge-Yukawa unified
model. The mass unit is TeV.}
\begin{center}
\begin{tabular}{|c|c||c|c|}
 \hline
$m_{\chi_1}$  &
0.22 &
$m_{\tilde{s}_1}=m_{\tilde{d}_1}$    &
1.18 \\ \hline
$m_{\chi_2}$  &
0.42 &
$m_{\tilde{s}_2}=m_{\tilde{d}_2}$     &
1.30
\\ \hline
$m_{\chi_3} $  &
0.90 &
$m_{\tilde{\tau}_1}$    &
0.42
\\ \hline
$m_{\chi_4}$  & 
0.91 &
$m_{\tilde{\tau}_2}$     &
0.59
\\ \hline
$m_{\chi^{\pm}_{1}} $  &
0.42 &
$m_{\tilde{\nu}_{\tau}}$ 
 & 
0.54
\\ \hline
$m_{\chi^{\pm}_{2}} $  & 0.91 &
$m_{\tilde{\mu}_1}=m_{\tilde{e}_1}$
 &
0.72
\\ \hline
$m_{\tilde{t}_1}$  &
0.87 &
$m_{\tilde{\mu}_2}=m_{\tilde{e}_2}$
 &
0.80
\\ \hline
$m_{\tilde{t}_2}$   &
1.03 & $m_{\tilde{\nu}_{\mu}}=m_{\tilde{\nu}_{e}}$ 
 & 0.72
\\ \hline
$m_{\tilde{b}_1}$   &
0.87  &$m_A$   & 0.33
\\  \hline
$m_{\tilde{b}_2}$   &
1.01  &$m_{H^{\pm}}$   & 0.34
\\  \hline
$m_{\tilde{c}_1}=m_{\tilde{u}_1}$   &
1.26  & $m_H$  & 0.33
\\  \hline
$m_{\tilde{c}_2}=m_{\tilde{u}_2}$   &
1.30  &$m_h$  & 0.124
\\  \hline
$M_3$   &
1.16 &  &
\\  \hline
\end{tabular}
\end{center}
\end{table}
The mass values 
\footnote{For the mass of the lightest
Higgs, the RG improved corrections \cite{haber2}
are included.} in Table 2 are the running masses
at $M_{\rm SUSY}$ which is $\sim  0.95 $ TeV 
 \footnote{$M_{\rm SUSY}$ is no longer an independent
parameter and we use $M_{\rm SUSY}^{2}=
(m_{\tilde{t}_1}^2+m_{\tilde{t}_2}^{2})/2$, where 
$m_{\tilde{t}_{1,2}}$ are the masses of the superpartners of the top
quark.} for $M=0.5$ TeV.
The prediction above depends basically only on 
the unified gaugino mass $M$, and
so the model has an extremely strong predictive power.
Note also that $~m_{H_u}^{2}~,~m_{H_d}^{2}$
and $\tan\beta$ (see the Higgs potential (\ref{potential1}) and
the definition (\ref{tanb})) are now fixed outside 
of the Higgs sector,
so that there is no guaranty that 
the Higgs potential (\ref{potential1}) yields the desired
 symmetry breaking
of $SU(2)_{\rm L}\times U(1)_{\rm Y}$ gauge symmetry.
Surprisingly, in the case at hand
 it does!  
(If the sign of $m_{H_u}^{2}$ in (\ref{softmass1}) 
were different, for instance, it would not do.) In Table 3 I give the
predictions  from the dimensionless sector of the model.
\begin{table}
\caption{The predictions 
from the dimensionless sector of the minimal model. 
($M=0.5$ TeV)}
\begin{center}
\begin{tabular}{|c|c|c|c|c|}
\hline
 $\alpha_{3}(M_Z)$ &
$\tan \beta$  &  $M_{\rm GUT}$ [GeV] 
 & $M_{b}$ [GeV]& $M_{t}$ [GeV]
\\ \hline
$0.119 $  &$48.8 $  & $1.47\times 10^{16}$
 & $5.4$  & 177\\ \hline
\end{tabular}
\end{center}
\end{table}
At last but not  least we would like to emphasize that the
reduction solutions (\ref{trilinear1}) and (\ref{softmass1})
do {\em not} lead to the flavor changing neural current (FCNC)
problem. This is not something put ad hoc by hand;
it is a consequence of the principle of reduction of couplings.

\section{Sum Rules for the Soft Supersymmetry-Breaking Parameters}
\subsection{Renormalization group invariant sum rules}
Now I would like to come the next topic.
To proceed I  recall the result
of the reduction of the SSB parameters in favor of the
unified gaugino mass $M$ in the minimal SUSY $SU(5)$ model
which I have discussed just above.
As we have seen, the reduction solutions for the
trilinear couplings are universal while those for the soft
scalar masses are not (see (\ref{trilinear1}) and \ref{softmass1})).
However,  if  one adds the soft scalar mass squared in an
appropriate way, one finds something interesting \cite{kkk1}.
For instance,
\be
M^2 &=& m_{t_L}^{2}+ m_{t_R}^{2}+ m_{H_u}^{2}=
 m_{b_L}^{2}+ m_{b_R}^{2}+ m_{H_d}^{2}~.
\label{sum1}
\ee
This is not an accidental coincidence.
One can in fact show that the sum rules in this form are 
RG invariant 
at one-loop \cite{kkk1}.

In last years there have been 
continues developments \cite{yamada1}--\cite{jack5} in
computing the RG functions in
softly broken supersymmetric Yang-Mills 
theories, and 
 the well-known  result on the QCD $\beta$-function obtained by
 Professor Zakharov and his 
 collaborators \cite{novikov1} \footnote{Klaus Sibold 
 pointed out that there is some correction
to this $\beta$-function. See
\cite{lucchesi1} for the argument.}
has been generalized 
so as to include to the SSB sector
\cite{yamada1}--\cite{jack5}, which is based 
on a clever spurion superfield
technique along with power 
counting \footnote{It is not clear at the moment 
in which 
class of renormalization schemes 
exactly the
result is  valid; a renormalization scheme independent
investigation of this result is certainly desirable.}.
Using this result, it is possible to find
a closed form of the sum rules that are RG invariant
to all orders in perturbation theory \cite{kazakov1}-\cite{jack5}.

To be specific, we consider a softly broken supersymmetric theory
described by the superpotential
\be
W &=&\frac{1}{6} \,Y^{ijk}\,\Phi_i \Phi_j \Phi_k
+\frac{1}{2} \,\mu^{ij}\,\Phi_i \Phi_j~,
\ee
along with the Lagrangian for the SSB terms,
\be
-{\cal L}_{\rm SB} &=&
\frac{1}{6} \,h^{ijk}\,\phi_i \phi_j \phi_k
+
\frac{1}{2} \,b^{ij}\,\phi_i \phi_j
+
\frac{1}{2} \,(m^2)^{j}_{i}\,\phi^{*\,i} \phi_j+
\frac{1}{2} \,M\,\lambda \lambda+\mbox{H.c.}~,
\ee
where $\Phi_i$ stands for a chiral superfield
with its scalar component $\phi_i$, and 
$\lambda$ is the gaugino field.
It has been found \cite{jack4} that the expressions
\footnote{The Yukawa couplings $Y^{ijk}$ 
and $\mu^{ij}$ are assumed to be
functions of the gauge coupling $g$.}
\be
b^{ij} &=& -M \mu^{ij}\frac{d \ln \mu^{ij}(g)}{d \ln g}~,\nn\\
h^{ijk} &=& -M \frac{d Y^{ijk}(g)}{d \ln g}~,
\label{trilinear2}\\
m^2_i &=&
\frac{1}{2}|M|^2 (g/\beta_g) \frac{d \gamma_i (g)}{d \ln g}~
\label{softmass2}
\ee
are RG invariant to all orders in perturbation theory
in a certain class of renormalization schemes,
which are the higher order results for the one-loop reduction
solutions
(\ref{trilinear1}) and (\ref{softmass1}). 
Similarly, the sum rule (\ref{sum1}) in 
higher orders becomes \cite{kkz1}
\be
m^2_i+m^2_j+m^2_k &=&
|M|^2 \{~
\frac{1}{1-g^2 C(G)/(8\pi^2)}\frac{d \ln Y^{ijk}}{d \ln g}
+\frac{1}{2}\frac{d^2 \ln Y^{ijk}}{d (\ln g)^2}~\}\nn\\
& &+\sum_\l
\frac{m^2_\l T(R_\l)}{C(G)-8\pi^2/g^2}
\frac{d \ln Y^{ijk}}{d \ln g}~,
\label{sum2}
\ee
in the renormalization scheme which 
corresponds to that of  \cite{novikov1}.
Here $C(G)$ is the 
quardratic Casimir in the adjoint representation,
$T(R)$ stands for the Dynkin index of the representation $R$,
$\beta_g$ is the $\beta$-function of the gauge coupling $g$,
$\gamma_i$ is the anomalous dimension of
$\Phi_i$. These
expressions look slightly complicated. But if one uses the freedom of
reparametrization \cite{oehme1}
(as discussed in the previous talk of Professor
Oehme), they can be transformed into a more simple 
form ($(d \ln Y^{ijk}/d \ln g)=1$):
\be
h^{ijk} &=&-Y^{ijk}(g) M~,
\label{trilinear3}\\
m^2_i+m^2_j+m^2_k &=&
|M|^2~
\frac{1}{1-g^2 C(G)/(8\pi^2)}+\sum_\l
\frac{m^2_\l T(R_\l)}{C(G)-8\pi^2/g^2}~,
\label{sum3}
\ee
It is exactly this form which coincides with the results
obtained in  certain orbifold models of 
superstrings \cite{kkmz1} 
\footnote{Tree-level sum rules (like (\ref{sum1})  in string theories
are found in \cite{kkk1}, \cite{BIM2}-\cite{ibanez1}}.
 I believe that this coincidence is not 
accidental, and I also
believe that
 target-space duality invariance  \cite{kikkawa1},
which is supposed to be
an exact symmetry of  compactified 
superstring theories \footnote{See \cite{polchinski},
for instance, for target-space duality.},
is most responsible for the coincidence. In fact there exist
already some indications for that.
I hope I can report on the true reason  of this interesting
coincidence in near future.

\subsection{Finiteness and sum rules}
At this stage it may be worthwhile to mention that the reduction solution
(\ref{trilinear2}) and the sum rules (\ref{sum2}) ensure
the finiteness of the SSB sector in a finite theory
\footnote{There exists a fine difference in the opinions about this 
point. See, for instance, \cite{kazakov1,jack4}.}.
For the $N=4$ supersymmetric Yang Mills theory
written in terms of $N=1$ superfields, for instance, we
have
$\sum_\l m^2_\l T(R_\l) =
(m^2_i+m^2_j+m^2_k) C(G)$ so that
the all order sum rule (\ref{sum3}) assumes the
tree level form $m^2_i+m^2_j+m^2_k=|M|^2$.
Applied to the finite $SU(5)$ model \cite{kkmz1} 
which I discussed in the
previous section (Table 1 presents the prediction from the dimensionless
sector), it means that 
the sum rules \cite{kkmz1} 
\be
m^{2}_{H_u}+
2  m^{2}_{{\bf 10}} &=&M^2~,~
m^{2}_{H_d}-2m^{2}_{{\bf 10}}=-\frac{M^2}{3}~,~
m^{2}_{\overline{{\bf 5}}}+
3m^{2}_{{\bf 10}}=\frac{4M^2}{3}
\ee
should be satisfied at and above $M_{\rm GUT}$
for the two-loop finiteness of the
SSB sector requires that, where
\be
m_{{\bf 10}} &=&
m_{ t_L}  = m_{ b_L}=
m_{ t_R}=m_{ \tau_R}~,~
m_{\overline{{\bf 5}}}
=m_{ b_R}  = 
m_{ \tau_L}=m_{ \nu_{\tau}}~.
\ee
In his casewe have an additional free parameter,
$m_{{\bf 10}}$, in the SSB sector.
It turned out that the mass of a superpartner of
the tau (s-tau) tends to become very light in this model.
Consequently, in order to obtain a neutral lightest
superparticle (LSP) (because we assume that $R$-
parity is intact), 
we have to have a large unified gravitino mass
$M \gsim 0.8$ TeV.
For $M= 1$ TeV, only the window
$0.62 ~~\mbox{TeV} < m_{{\bf 10}} <0.66 ~~\mbox{TeV}$ 
is allowed.
In Table 4 we give the prediction of the superpartner
spectrum of the model for $ m_{{\bf 10}}=0.62/0.66$ TeV
and $M =1$ TeV. We have assumed the universal 
soft masses for the first two generations.
But this assumption does not change practically
our prediction of the spectrum expect for those that
are directly of the first two generations.
\begin{table}
\caption{The predictions of the superpartner spectrum
for the finite $SU(5)$ model.
$M=1$ TeV and $m_{\bf 10}=0.62/0.66$ TeV.}
\begin{center}
\begin{tabular}{|c|c||c|c|}
 \hline
$m_{\chi_1}$  &
0.45/0.45 &
$m_{\tilde{s}_1}=m_{\tilde{d}_1}$    &
1.95/1.95 \\ \hline
$m_{\chi_2}$  &
0.84/0.84 &
$m_{\tilde{s}_2}=m_{\tilde{d}_2}$     &
2.06/2.05
\\ \hline
$m_{\chi_3} $  &
1.29/1.32 &
$m_{\tilde{\tau}_1}$    &
0.46/0.46
\\ \hline
$m_{\chi_4}$  & 
1.29/1.32 &
$m_{\tilde{\tau}_2}$     &
0.73/0.66
\\ \hline
$m_{\chi^{\pm}_{1}} $  &
0.84/0.84 &
$m_{\tilde{\nu}_{\tau}}$ 
 & 
0.70/0.57
\\ \hline
$m_{\chi^{\pm}_{2}} $  & 1.29/1.32 &
$m_{\tilde{\mu}_1}=m_{\tilde{e}_1}$
 &
0.70/0.71
\\ \hline
$m_{\tilde{t}_1}$  &
1.50/1.51 &
$m_{\tilde{\mu}_2}=m_{\tilde{e}_2}$
 &
0.89/0.89
\\ \hline
$m_{\tilde{t}_2}$   &
1.72/1.74 & $m_{\tilde{\nu}_{\mu}}=m_{\tilde{\nu}_{e}}$ 
 & 0.89/0.88
\\ \hline
$m_{\tilde{b}_1}$   &
1.51/1.46  &$m_A$   & 0.63/0.77
\\  \hline
$m_{\tilde{b}_2}$   &
1.70/1.71  &$m_{H^{\pm}}$   & 0.63/0.77
\\  \hline
$m_{\tilde{c}_1}=m_{\tilde{u}_1}$   &
1.96/1.96  & $m_H$  & 0.63/0.77
\\  \hline
$m_{\tilde{c}_2}=m_{\tilde{u}_2}$   &
2.05/2.05  &$m_h$  & 0.127/0.127
\\  \hline
$M_3$   &
2.21/2.21  & &
\\  \hline
\end{tabular}
\end{center}
\end{table}
 
\subsection{Sum rules in the superpartner spectrum}
The sum rules (\ref{sum1}) or (\ref{sum3}) can be translated into
the sum rules of the superpartner spectrum of the MSSM \cite{kkk2}
as I will show now.
To be specific we assume
an $SU(5)$ type Gauge-Yukawa unification 
in the third generation of the form
(\ref{gyu2}). For a given model, 
the constants $\kappa$'s are fixed, but here we consider
them as free parameters. As before we use the tau mass $M_{\tau}$
as an input parameter, and we go from the
parameter space $(\kappa_t~,~\kappa_b)$ to 
another  one $(\kappa_t~,~\tan\beta)$, because in this analysis
we use 
 the physical top quark $M_{t}$, too,  as an input parameter.
Then the unification conditions of 
the gauge and Yukawa couplings of the MSSM (i.e.,
$g=g_1=g_2=g_3~,~g_b=g_{\tau}$)
fixes the allowed region (line) in the $\kappa_t - \tan\beta$ space
for a given value of the unified gaugino mass $M$.
The parameter space in the SSB sector  at  $M_{\rm GUT}$
is  constrained  due
to unification:
\begin{eqnarray}
M &=& M_1= M_2 =M_3 ~,\nn\\
m_{t_R}^{2} &=&m_{t_L}^{2}=
m_{b_L}^{2}= m_{\tau_R}^{2}~,~
m_{b_R}^{2}= m_{\tau_L}^{2}=m_{\nu_{\tau}}^{2}~,
\ee
where $M_i~(i=1,2,3)$ are the gaugino masses for 
$U(1)_{\rm Y}$ (bino), $SU(2)_{\rm L}$ (wino) and $SU(3)_{\rm C}$ (gluino).
And the one-loop sum rules at  $M_{\rm GUT}$ yield
\be
h_t &=&-M~,
~h_b =h_{\tau}=-M\, g_b~,~
M^2=m_{\Sigma (t)}^{2}=
m_{\Sigma (b)}^{2}=m_{\Sigma (\tau)}^{2}~,
\label{sum5}
\ee
where
\be
m_{\Sigma (t)}^{2}&\equiv& m_{t_R}^{2}+
m_{t_L}^{2}+m_{H_u}^{2}~,~
m_{\Sigma (b,\tau)}^{2}\equiv 
m_{b_R,\tau_R}^{2}+
m_{b_L,\tau_L}^{2}+m_{H_d}^{2}~.
\end{eqnarray}
(The above equations are the same as 
(\ref{trilinear1}) and (\ref{sum1}), respectively.)
I would like to emphasize that in the one-loop RG evolution of
$m_{\Sigma}^{2}$'s 
in the MSSM only  the same combinations 
of the sum of $m_{i}^{2}$'s
enter. Therefore, as far as we are interested in
the evolution of $m_{\Sigma}^{2}$'s,
we have only one additional parameter
$M_{\rm SUSY}$.
To derive the announced sum rules for the superpartner spectrum,
we  define
\begin{eqnarray}
s_i &\equiv& m^2_{\Sigma(i)}/M_3^2~~(i=t,b,\tau)~~\mbox{at}~~
Q=M_{\rm SUSY}.
\label{sbt}
\end{eqnarray}
The parameters $s_i$'s do not depend on the value of the unified gaugino
mass $M$,
but they do on $\tan \beta$.
This dependence is shown in Fig. 4.
We then  express the  masses of the superpartners
in terms of the
soft scalar masses and the masses of the
ordinary particles to obtain
the sum rules \cite{kkk2},
\begin{eqnarray}
-\cos 2\beta ~m_{A}^{2}
&=&(s_{b}-s_{t} )M_{3}^{2} +2 (\hat{m}_{t}^{2}-
m_{t}^{2}) 
-2 (\hat{m}_{b}^{2}-
m_{b}^{2}) \nonumber\\
&=& (s_{\tau}-s_{t}) M_{3}^{2} +2 (\hat{m}_{t}^{2}-m_{t}^{2})
-2 (\hat{m}_{\tau}^{2}-m_{\tau}^{2}),
\label{sum4}
\end{eqnarray}
where $m_{A}^{2}$ is the neutral pseudoscalar Higgs
mass squared, and $\hat{m}_{i}^{2}$
stands for the arithmetic mean of the two
corresponding scalar superparticle mass squared. 
\begin{center}
\setlength{\unitlength}{0.240900pt}
\ifx\plotpoint\undefined\newsavebox{\plotpoint}\fi
\sbox{\plotpoint}{\rule[-0.200pt]{0.400pt}{0.400pt}}%
\begin{picture}(1500,900)(0,0)
\font\gnuplot=cmr10 at 10pt
\gnuplot
\sbox{\plotpoint}{\rule[-0.200pt]{0.400pt}{0.400pt}}%
\put(220.0,189.0){\rule[-0.200pt]{292.934pt}{0.400pt}}
\put(220.0,113.0){\rule[-0.200pt]{0.400pt}{184.048pt}}
\put(220.0,113.0){\rule[-0.200pt]{4.818pt}{0.400pt}}
\put(198,113){\makebox(0,0)[r]{-0.2}}
\put(1416.0,113.0){\rule[-0.200pt]{4.818pt}{0.400pt}}
\put(220.0,189.0){\rule[-0.200pt]{4.818pt}{0.400pt}}
\put(198,189){\makebox(0,0)[r]{0}}
\put(1416.0,189.0){\rule[-0.200pt]{4.818pt}{0.400pt}}
\put(220.0,266.0){\rule[-0.200pt]{4.818pt}{0.400pt}}
\put(198,266){\makebox(0,0)[r]{0.2}}
\put(1416.0,266.0){\rule[-0.200pt]{4.818pt}{0.400pt}}
\put(220.0,342.0){\rule[-0.200pt]{4.818pt}{0.400pt}}
\put(198,342){\makebox(0,0)[r]{0.4}}
\put(1416.0,342.0){\rule[-0.200pt]{4.818pt}{0.400pt}}
\put(220.0,419.0){\rule[-0.200pt]{4.818pt}{0.400pt}}
\put(198,419){\makebox(0,0)[r]{0.6}}
\put(1416.0,419.0){\rule[-0.200pt]{4.818pt}{0.400pt}}
\put(220.0,495.0){\rule[-0.200pt]{4.818pt}{0.400pt}}
\put(198,495){\makebox(0,0)[r]{0.8}}
\put(1416.0,495.0){\rule[-0.200pt]{4.818pt}{0.400pt}}
\put(220.0,571.0){\rule[-0.200pt]{4.818pt}{0.400pt}}
\put(198,571){\makebox(0,0)[r]{1}}
\put(1416.0,571.0){\rule[-0.200pt]{4.818pt}{0.400pt}}
\put(220.0,648.0){\rule[-0.200pt]{4.818pt}{0.400pt}}
\put(198,648){\makebox(0,0)[r]{1.2}}
\put(1416.0,648.0){\rule[-0.200pt]{4.818pt}{0.400pt}}
\put(220.0,724.0){\rule[-0.200pt]{4.818pt}{0.400pt}}
\put(198,724){\makebox(0,0)[r]{1.4}}
\put(1416.0,724.0){\rule[-0.200pt]{4.818pt}{0.400pt}}
\put(220.0,801.0){\rule[-0.200pt]{4.818pt}{0.400pt}}
\put(198,801){\makebox(0,0)[r]{1.6}}
\put(1416.0,801.0){\rule[-0.200pt]{4.818pt}{0.400pt}}
\put(220.0,877.0){\rule[-0.200pt]{4.818pt}{0.400pt}}
\put(198,877){\makebox(0,0)[r]{1.8}}
\put(1416.0,877.0){\rule[-0.200pt]{4.818pt}{0.400pt}}
\put(220.0,113.0){\rule[-0.200pt]{0.400pt}{4.818pt}}
\put(220,68){\makebox(0,0){0}}
\put(220.0,857.0){\rule[-0.200pt]{0.400pt}{4.818pt}}
\put(342.0,113.0){\rule[-0.200pt]{0.400pt}{4.818pt}}
\put(342,68){\makebox(0,0){5}}
\put(342.0,857.0){\rule[-0.200pt]{0.400pt}{4.818pt}}
\put(463.0,113.0){\rule[-0.200pt]{0.400pt}{4.818pt}}
\put(463,68){\makebox(0,0){10}}
\put(463.0,857.0){\rule[-0.200pt]{0.400pt}{4.818pt}}
\put(585.0,113.0){\rule[-0.200pt]{0.400pt}{4.818pt}}
\put(585,68){\makebox(0,0){15}}
\put(585.0,857.0){\rule[-0.200pt]{0.400pt}{4.818pt}}
\put(706.0,113.0){\rule[-0.200pt]{0.400pt}{4.818pt}}
\put(706,68){\makebox(0,0){20}}
\put(706.0,857.0){\rule[-0.200pt]{0.400pt}{4.818pt}}
\put(828.0,113.0){\rule[-0.200pt]{0.400pt}{4.818pt}}
\put(828,68){\makebox(0,0){25}}
\put(828.0,857.0){\rule[-0.200pt]{0.400pt}{4.818pt}}
\put(950.0,113.0){\rule[-0.200pt]{0.400pt}{4.818pt}}
\put(950,68){\makebox(0,0){30}}
\put(950.0,857.0){\rule[-0.200pt]{0.400pt}{4.818pt}}
\put(1071.0,113.0){\rule[-0.200pt]{0.400pt}{4.818pt}}
\put(1071,68){\makebox(0,0){35}}
\put(1071.0,857.0){\rule[-0.200pt]{0.400pt}{4.818pt}}
\put(1193.0,113.0){\rule[-0.200pt]{0.400pt}{4.818pt}}
\put(1193,68){\makebox(0,0){40}}
\put(1193.0,857.0){\rule[-0.200pt]{0.400pt}{4.818pt}}
\put(1314.0,113.0){\rule[-0.200pt]{0.400pt}{4.818pt}}
\put(1314,68){\makebox(0,0){45}}
\put(1314.0,857.0){\rule[-0.200pt]{0.400pt}{4.818pt}}
\put(1436.0,113.0){\rule[-0.200pt]{0.400pt}{4.818pt}}
\put(1436,68){\makebox(0,0){50}}
\put(1436.0,857.0){\rule[-0.200pt]{0.400pt}{4.818pt}}
\put(220.0,113.0){\rule[-0.200pt]{292.934pt}{0.400pt}}
\put(1436.0,113.0){\rule[-0.200pt]{0.400pt}{184.048pt}}
\put(220.0,877.0){\rule[-0.200pt]{292.934pt}{0.400pt}}
\put(45,495){\makebox(0,0){$s_{t,b,\tau}$}}
\put(828,23){\makebox(0,0){$\tan \beta$}}
\put(1071,724){\makebox(0,0)[r]{$b$}}
\put(828,571){\makebox(0,0)[r]{$t$}}
\put(950,342){\makebox(0,0)[r]{$\tau$}}
\put(220.0,113.0){\rule[-0.200pt]{0.400pt}{184.048pt}}
\put(269,464){\usebox{\plotpoint}}
\multiput(269.00,464.59)(1.368,0.489){15}{\rule{1.167pt}{0.118pt}}
\multiput(269.00,463.17)(21.579,9.000){2}{\rule{0.583pt}{0.400pt}}
\multiput(293.00,473.59)(2.602,0.477){7}{\rule{2.020pt}{0.115pt}}
\multiput(293.00,472.17)(19.807,5.000){2}{\rule{1.010pt}{0.400pt}}
\multiput(317.00,478.61)(5.374,0.447){3}{\rule{3.433pt}{0.108pt}}
\multiput(317.00,477.17)(17.874,3.000){2}{\rule{1.717pt}{0.400pt}}
\put(342,481.17){\rule{4.900pt}{0.400pt}}
\multiput(342.00,480.17)(13.830,2.000){2}{\rule{2.450pt}{0.400pt}}
\put(390,482.67){\rule{6.023pt}{0.400pt}}
\multiput(390.00,482.17)(12.500,1.000){2}{\rule{3.011pt}{0.400pt}}
\put(366.0,483.0){\rule[-0.200pt]{5.782pt}{0.400pt}}
\put(463,482.67){\rule{6.023pt}{0.400pt}}
\multiput(463.00,483.17)(12.500,-1.000){2}{\rule{3.011pt}{0.400pt}}
\put(415.0,484.0){\rule[-0.200pt]{11.563pt}{0.400pt}}
\put(536,481.67){\rule{5.782pt}{0.400pt}}
\multiput(536.00,482.17)(12.000,-1.000){2}{\rule{2.891pt}{0.400pt}}
\put(488.0,483.0){\rule[-0.200pt]{11.563pt}{0.400pt}}
\put(585,480.67){\rule{5.782pt}{0.400pt}}
\multiput(585.00,481.17)(12.000,-1.000){2}{\rule{2.891pt}{0.400pt}}
\put(609,479.67){\rule{5.782pt}{0.400pt}}
\multiput(609.00,480.17)(12.000,-1.000){2}{\rule{2.891pt}{0.400pt}}
\put(560.0,482.0){\rule[-0.200pt]{6.022pt}{0.400pt}}
\put(658,478.67){\rule{5.782pt}{0.400pt}}
\multiput(658.00,479.17)(12.000,-1.000){2}{\rule{2.891pt}{0.400pt}}
\put(682,477.67){\rule{5.782pt}{0.400pt}}
\multiput(682.00,478.17)(12.000,-1.000){2}{\rule{2.891pt}{0.400pt}}
\put(633.0,480.0){\rule[-0.200pt]{6.022pt}{0.400pt}}
\put(731,476.67){\rule{5.782pt}{0.400pt}}
\multiput(731.00,477.17)(12.000,-1.000){2}{\rule{2.891pt}{0.400pt}}
\put(755,475.67){\rule{5.782pt}{0.400pt}}
\multiput(755.00,476.17)(12.000,-1.000){2}{\rule{2.891pt}{0.400pt}}
\put(779,474.67){\rule{6.023pt}{0.400pt}}
\multiput(779.00,475.17)(12.500,-1.000){2}{\rule{3.011pt}{0.400pt}}
\put(706.0,478.0){\rule[-0.200pt]{6.022pt}{0.400pt}}
\put(828,473.67){\rule{5.782pt}{0.400pt}}
\multiput(828.00,474.17)(12.000,-1.000){2}{\rule{2.891pt}{0.400pt}}
\put(852,472.67){\rule{6.023pt}{0.400pt}}
\multiput(852.00,473.17)(12.500,-1.000){2}{\rule{3.011pt}{0.400pt}}
\put(804.0,475.0){\rule[-0.200pt]{5.782pt}{0.400pt}}
\put(901,471.67){\rule{5.782pt}{0.400pt}}
\multiput(901.00,472.17)(12.000,-1.000){2}{\rule{2.891pt}{0.400pt}}
\put(925,470.67){\rule{6.023pt}{0.400pt}}
\multiput(925.00,471.17)(12.500,-1.000){2}{\rule{3.011pt}{0.400pt}}
\put(877.0,473.0){\rule[-0.200pt]{5.782pt}{0.400pt}}
\put(974,469.67){\rule{5.782pt}{0.400pt}}
\multiput(974.00,470.17)(12.000,-1.000){2}{\rule{2.891pt}{0.400pt}}
\put(950.0,471.0){\rule[-0.200pt]{5.782pt}{0.400pt}}
\put(1023,468.67){\rule{5.782pt}{0.400pt}}
\multiput(1023.00,469.17)(12.000,-1.000){2}{\rule{2.891pt}{0.400pt}}
\put(998.0,470.0){\rule[-0.200pt]{6.022pt}{0.400pt}}
\put(1071,467.67){\rule{6.023pt}{0.400pt}}
\multiput(1071.00,468.17)(12.500,-1.000){2}{\rule{3.011pt}{0.400pt}}
\put(1047.0,469.0){\rule[-0.200pt]{5.782pt}{0.400pt}}
\put(1144,466.67){\rule{5.782pt}{0.400pt}}
\multiput(1144.00,467.17)(12.000,-1.000){2}{\rule{2.891pt}{0.400pt}}
\put(1096.0,468.0){\rule[-0.200pt]{11.563pt}{0.400pt}}
\put(1387,466.67){\rule{6.023pt}{0.400pt}}
\multiput(1387.00,466.17)(12.500,1.000){2}{\rule{3.011pt}{0.400pt}}
\put(1168.0,467.0){\rule[-0.200pt]{52.757pt}{0.400pt}}
\put(1412.0,468.0){\rule[-0.200pt]{5.782pt}{0.400pt}}
\put(269,809){\usebox{\plotpoint}}
\put(317,807.17){\rule{5.100pt}{0.400pt}}
\multiput(317.00,808.17)(14.415,-2.000){2}{\rule{2.550pt}{0.400pt}}
\multiput(342.00,805.95)(5.151,-0.447){3}{\rule{3.300pt}{0.108pt}}
\multiput(342.00,806.17)(17.151,-3.000){2}{\rule{1.650pt}{0.400pt}}
\multiput(366.00,802.95)(5.151,-0.447){3}{\rule{3.300pt}{0.108pt}}
\multiput(366.00,803.17)(17.151,-3.000){2}{\rule{1.650pt}{0.400pt}}
\multiput(390.00,799.94)(3.552,-0.468){5}{\rule{2.600pt}{0.113pt}}
\multiput(390.00,800.17)(19.604,-4.000){2}{\rule{1.300pt}{0.400pt}}
\multiput(415.00,795.94)(3.406,-0.468){5}{\rule{2.500pt}{0.113pt}}
\multiput(415.00,796.17)(18.811,-4.000){2}{\rule{1.250pt}{0.400pt}}
\multiput(439.00,791.93)(2.602,-0.477){7}{\rule{2.020pt}{0.115pt}}
\multiput(439.00,792.17)(19.807,-5.000){2}{\rule{1.010pt}{0.400pt}}
\multiput(463.00,786.93)(2.714,-0.477){7}{\rule{2.100pt}{0.115pt}}
\multiput(463.00,787.17)(20.641,-5.000){2}{\rule{1.050pt}{0.400pt}}
\multiput(488.00,781.93)(2.118,-0.482){9}{\rule{1.700pt}{0.116pt}}
\multiput(488.00,782.17)(20.472,-6.000){2}{\rule{0.850pt}{0.400pt}}
\multiput(512.00,775.93)(2.118,-0.482){9}{\rule{1.700pt}{0.116pt}}
\multiput(512.00,776.17)(20.472,-6.000){2}{\rule{0.850pt}{0.400pt}}
\multiput(536.00,769.93)(2.118,-0.482){9}{\rule{1.700pt}{0.116pt}}
\multiput(536.00,770.17)(20.472,-6.000){2}{\rule{0.850pt}{0.400pt}}
\multiput(560.00,763.93)(1.865,-0.485){11}{\rule{1.529pt}{0.117pt}}
\multiput(560.00,764.17)(21.827,-7.000){2}{\rule{0.764pt}{0.400pt}}
\multiput(585.00,756.93)(1.789,-0.485){11}{\rule{1.471pt}{0.117pt}}
\multiput(585.00,757.17)(20.946,-7.000){2}{\rule{0.736pt}{0.400pt}}
\multiput(609.00,749.93)(1.550,-0.488){13}{\rule{1.300pt}{0.117pt}}
\multiput(609.00,750.17)(21.302,-8.000){2}{\rule{0.650pt}{0.400pt}}
\multiput(633.00,741.93)(1.865,-0.485){11}{\rule{1.529pt}{0.117pt}}
\multiput(633.00,742.17)(21.827,-7.000){2}{\rule{0.764pt}{0.400pt}}
\multiput(658.00,734.93)(1.550,-0.488){13}{\rule{1.300pt}{0.117pt}}
\multiput(658.00,735.17)(21.302,-8.000){2}{\rule{0.650pt}{0.400pt}}
\multiput(682.00,726.93)(1.368,-0.489){15}{\rule{1.167pt}{0.118pt}}
\multiput(682.00,727.17)(21.579,-9.000){2}{\rule{0.583pt}{0.400pt}}
\multiput(706.00,717.93)(1.616,-0.488){13}{\rule{1.350pt}{0.117pt}}
\multiput(706.00,718.17)(22.198,-8.000){2}{\rule{0.675pt}{0.400pt}}
\multiput(731.00,709.93)(1.368,-0.489){15}{\rule{1.167pt}{0.118pt}}
\multiput(731.00,710.17)(21.579,-9.000){2}{\rule{0.583pt}{0.400pt}}
\multiput(755.00,700.93)(1.368,-0.489){15}{\rule{1.167pt}{0.118pt}}
\multiput(755.00,701.17)(21.579,-9.000){2}{\rule{0.583pt}{0.400pt}}
\multiput(779.00,691.93)(1.427,-0.489){15}{\rule{1.211pt}{0.118pt}}
\multiput(779.00,692.17)(22.486,-9.000){2}{\rule{0.606pt}{0.400pt}}
\multiput(804.00,682.93)(1.368,-0.489){15}{\rule{1.167pt}{0.118pt}}
\multiput(804.00,683.17)(21.579,-9.000){2}{\rule{0.583pt}{0.400pt}}
\multiput(828.00,673.93)(1.368,-0.489){15}{\rule{1.167pt}{0.118pt}}
\multiput(828.00,674.17)(21.579,-9.000){2}{\rule{0.583pt}{0.400pt}}
\multiput(852.00,664.93)(1.427,-0.489){15}{\rule{1.211pt}{0.118pt}}
\multiput(852.00,665.17)(22.486,-9.000){2}{\rule{0.606pt}{0.400pt}}
\multiput(877.00,655.93)(1.368,-0.489){15}{\rule{1.167pt}{0.118pt}}
\multiput(877.00,656.17)(21.579,-9.000){2}{\rule{0.583pt}{0.400pt}}
\multiput(901.00,646.92)(1.225,-0.491){17}{\rule{1.060pt}{0.118pt}}
\multiput(901.00,647.17)(21.800,-10.000){2}{\rule{0.530pt}{0.400pt}}
\multiput(925.00,636.93)(1.427,-0.489){15}{\rule{1.211pt}{0.118pt}}
\multiput(925.00,637.17)(22.486,-9.000){2}{\rule{0.606pt}{0.400pt}}
\multiput(950.00,627.93)(1.368,-0.489){15}{\rule{1.167pt}{0.118pt}}
\multiput(950.00,628.17)(21.579,-9.000){2}{\rule{0.583pt}{0.400pt}}
\multiput(974.00,618.93)(1.368,-0.489){15}{\rule{1.167pt}{0.118pt}}
\multiput(974.00,619.17)(21.579,-9.000){2}{\rule{0.583pt}{0.400pt}}
\multiput(998.00,609.93)(1.427,-0.489){15}{\rule{1.211pt}{0.118pt}}
\multiput(998.00,610.17)(22.486,-9.000){2}{\rule{0.606pt}{0.400pt}}
\multiput(1023.00,600.93)(1.550,-0.488){13}{\rule{1.300pt}{0.117pt}}
\multiput(1023.00,601.17)(21.302,-8.000){2}{\rule{0.650pt}{0.400pt}}
\multiput(1047.00,592.93)(1.368,-0.489){15}{\rule{1.167pt}{0.118pt}}
\multiput(1047.00,593.17)(21.579,-9.000){2}{\rule{0.583pt}{0.400pt}}
\multiput(1071.00,583.93)(1.616,-0.488){13}{\rule{1.350pt}{0.117pt}}
\multiput(1071.00,584.17)(22.198,-8.000){2}{\rule{0.675pt}{0.400pt}}
\multiput(1096.00,575.93)(1.550,-0.488){13}{\rule{1.300pt}{0.117pt}}
\multiput(1096.00,576.17)(21.302,-8.000){2}{\rule{0.650pt}{0.400pt}}
\multiput(1120.00,567.93)(1.550,-0.488){13}{\rule{1.300pt}{0.117pt}}
\multiput(1120.00,568.17)(21.302,-8.000){2}{\rule{0.650pt}{0.400pt}}
\multiput(1144.00,559.93)(1.550,-0.488){13}{\rule{1.300pt}{0.117pt}}
\multiput(1144.00,560.17)(21.302,-8.000){2}{\rule{0.650pt}{0.400pt}}
\multiput(1168.00,551.93)(1.865,-0.485){11}{\rule{1.529pt}{0.117pt}}
\multiput(1168.00,552.17)(21.827,-7.000){2}{\rule{0.764pt}{0.400pt}}
\multiput(1193.00,544.93)(1.789,-0.485){11}{\rule{1.471pt}{0.117pt}}
\multiput(1193.00,545.17)(20.946,-7.000){2}{\rule{0.736pt}{0.400pt}}
\multiput(1217.00,537.93)(1.789,-0.485){11}{\rule{1.471pt}{0.117pt}}
\multiput(1217.00,538.17)(20.946,-7.000){2}{\rule{0.736pt}{0.400pt}}
\multiput(1241.00,530.93)(1.865,-0.485){11}{\rule{1.529pt}{0.117pt}}
\multiput(1241.00,531.17)(21.827,-7.000){2}{\rule{0.764pt}{0.400pt}}
\multiput(1266.00,523.93)(2.118,-0.482){9}{\rule{1.700pt}{0.116pt}}
\multiput(1266.00,524.17)(20.472,-6.000){2}{\rule{0.850pt}{0.400pt}}
\multiput(1290.00,517.93)(2.602,-0.477){7}{\rule{2.020pt}{0.115pt}}
\multiput(1290.00,518.17)(19.807,-5.000){2}{\rule{1.010pt}{0.400pt}}
\multiput(1314.00,512.93)(2.208,-0.482){9}{\rule{1.767pt}{0.116pt}}
\multiput(1314.00,513.17)(21.333,-6.000){2}{\rule{0.883pt}{0.400pt}}
\multiput(1339.00,506.93)(2.602,-0.477){7}{\rule{2.020pt}{0.115pt}}
\multiput(1339.00,507.17)(19.807,-5.000){2}{\rule{1.010pt}{0.400pt}}
\multiput(1363.00,501.94)(3.406,-0.468){5}{\rule{2.500pt}{0.113pt}}
\multiput(1363.00,502.17)(18.811,-4.000){2}{\rule{1.250pt}{0.400pt}}
\multiput(1387.00,497.93)(2.714,-0.477){7}{\rule{2.100pt}{0.115pt}}
\multiput(1387.00,498.17)(20.641,-5.000){2}{\rule{1.050pt}{0.400pt}}
\multiput(1412.00,492.95)(5.151,-0.447){3}{\rule{3.300pt}{0.108pt}}
\multiput(1412.00,493.17)(17.151,-3.000){2}{\rule{1.650pt}{0.400pt}}
\put(269.0,809.0){\rule[-0.200pt]{11.563pt}{0.400pt}}
\put(269,320){\usebox{\plotpoint}}
\put(269,318.67){\rule{5.782pt}{0.400pt}}
\multiput(269.00,319.17)(12.000,-1.000){2}{\rule{2.891pt}{0.400pt}}
\put(293,317.17){\rule{4.900pt}{0.400pt}}
\multiput(293.00,318.17)(13.830,-2.000){2}{\rule{2.450pt}{0.400pt}}
\put(317,315.67){\rule{6.023pt}{0.400pt}}
\multiput(317.00,316.17)(12.500,-1.000){2}{\rule{3.011pt}{0.400pt}}
\put(342,314.17){\rule{4.900pt}{0.400pt}}
\multiput(342.00,315.17)(13.830,-2.000){2}{\rule{2.450pt}{0.400pt}}
\put(366,312.17){\rule{4.900pt}{0.400pt}}
\multiput(366.00,313.17)(13.830,-2.000){2}{\rule{2.450pt}{0.400pt}}
\put(390,310.17){\rule{5.100pt}{0.400pt}}
\multiput(390.00,311.17)(14.415,-2.000){2}{\rule{2.550pt}{0.400pt}}
\multiput(415.00,308.95)(5.151,-0.447){3}{\rule{3.300pt}{0.108pt}}
\multiput(415.00,309.17)(17.151,-3.000){2}{\rule{1.650pt}{0.400pt}}
\put(439,305.17){\rule{4.900pt}{0.400pt}}
\multiput(439.00,306.17)(13.830,-2.000){2}{\rule{2.450pt}{0.400pt}}
\multiput(463.00,303.94)(3.552,-0.468){5}{\rule{2.600pt}{0.113pt}}
\multiput(463.00,304.17)(19.604,-4.000){2}{\rule{1.300pt}{0.400pt}}
\multiput(488.00,299.95)(5.151,-0.447){3}{\rule{3.300pt}{0.108pt}}
\multiput(488.00,300.17)(17.151,-3.000){2}{\rule{1.650pt}{0.400pt}}
\multiput(512.00,296.94)(3.406,-0.468){5}{\rule{2.500pt}{0.113pt}}
\multiput(512.00,297.17)(18.811,-4.000){2}{\rule{1.250pt}{0.400pt}}
\multiput(536.00,292.95)(5.151,-0.447){3}{\rule{3.300pt}{0.108pt}}
\multiput(536.00,293.17)(17.151,-3.000){2}{\rule{1.650pt}{0.400pt}}
\multiput(560.00,289.94)(3.552,-0.468){5}{\rule{2.600pt}{0.113pt}}
\multiput(560.00,290.17)(19.604,-4.000){2}{\rule{1.300pt}{0.400pt}}
\multiput(585.00,285.93)(2.602,-0.477){7}{\rule{2.020pt}{0.115pt}}
\multiput(585.00,286.17)(19.807,-5.000){2}{\rule{1.010pt}{0.400pt}}
\multiput(609.00,280.94)(3.406,-0.468){5}{\rule{2.500pt}{0.113pt}}
\multiput(609.00,281.17)(18.811,-4.000){2}{\rule{1.250pt}{0.400pt}}
\multiput(633.00,276.93)(2.714,-0.477){7}{\rule{2.100pt}{0.115pt}}
\multiput(633.00,277.17)(20.641,-5.000){2}{\rule{1.050pt}{0.400pt}}
\multiput(658.00,271.93)(2.602,-0.477){7}{\rule{2.020pt}{0.115pt}}
\multiput(658.00,272.17)(19.807,-5.000){2}{\rule{1.010pt}{0.400pt}}
\multiput(682.00,266.93)(2.602,-0.477){7}{\rule{2.020pt}{0.115pt}}
\multiput(682.00,267.17)(19.807,-5.000){2}{\rule{1.010pt}{0.400pt}}
\multiput(706.00,261.93)(2.714,-0.477){7}{\rule{2.100pt}{0.115pt}}
\multiput(706.00,262.17)(20.641,-5.000){2}{\rule{1.050pt}{0.400pt}}
\multiput(731.00,256.93)(2.602,-0.477){7}{\rule{2.020pt}{0.115pt}}
\multiput(731.00,257.17)(19.807,-5.000){2}{\rule{1.010pt}{0.400pt}}
\multiput(755.00,251.93)(2.602,-0.477){7}{\rule{2.020pt}{0.115pt}}
\multiput(755.00,252.17)(19.807,-5.000){2}{\rule{1.010pt}{0.400pt}}
\multiput(779.00,246.93)(2.714,-0.477){7}{\rule{2.100pt}{0.115pt}}
\multiput(779.00,247.17)(20.641,-5.000){2}{\rule{1.050pt}{0.400pt}}
\multiput(804.00,241.93)(2.118,-0.482){9}{\rule{1.700pt}{0.116pt}}
\multiput(804.00,242.17)(20.472,-6.000){2}{\rule{0.850pt}{0.400pt}}
\multiput(828.00,235.93)(2.602,-0.477){7}{\rule{2.020pt}{0.115pt}}
\multiput(828.00,236.17)(19.807,-5.000){2}{\rule{1.010pt}{0.400pt}}
\multiput(852.00,230.93)(2.208,-0.482){9}{\rule{1.767pt}{0.116pt}}
\multiput(852.00,231.17)(21.333,-6.000){2}{\rule{0.883pt}{0.400pt}}
\multiput(877.00,224.93)(2.602,-0.477){7}{\rule{2.020pt}{0.115pt}}
\multiput(877.00,225.17)(19.807,-5.000){2}{\rule{1.010pt}{0.400pt}}
\multiput(901.00,219.93)(2.118,-0.482){9}{\rule{1.700pt}{0.116pt}}
\multiput(901.00,220.17)(20.472,-6.000){2}{\rule{0.850pt}{0.400pt}}
\multiput(925.00,213.93)(2.714,-0.477){7}{\rule{2.100pt}{0.115pt}}
\multiput(925.00,214.17)(20.641,-5.000){2}{\rule{1.050pt}{0.400pt}}
\multiput(950.00,208.93)(2.118,-0.482){9}{\rule{1.700pt}{0.116pt}}
\multiput(950.00,209.17)(20.472,-6.000){2}{\rule{0.850pt}{0.400pt}}
\multiput(974.00,202.93)(2.602,-0.477){7}{\rule{2.020pt}{0.115pt}}
\multiput(974.00,203.17)(19.807,-5.000){2}{\rule{1.010pt}{0.400pt}}
\multiput(998.00,197.93)(2.208,-0.482){9}{\rule{1.767pt}{0.116pt}}
\multiput(998.00,198.17)(21.333,-6.000){2}{\rule{0.883pt}{0.400pt}}
\multiput(1023.00,191.93)(2.602,-0.477){7}{\rule{2.020pt}{0.115pt}}
\multiput(1023.00,192.17)(19.807,-5.000){2}{\rule{1.010pt}{0.400pt}}
\multiput(1047.00,186.93)(2.602,-0.477){7}{\rule{2.020pt}{0.115pt}}
\multiput(1047.00,187.17)(19.807,-5.000){2}{\rule{1.010pt}{0.400pt}}
\multiput(1071.00,181.93)(2.714,-0.477){7}{\rule{2.100pt}{0.115pt}}
\multiput(1071.00,182.17)(20.641,-5.000){2}{\rule{1.050pt}{0.400pt}}
\multiput(1096.00,176.93)(2.602,-0.477){7}{\rule{2.020pt}{0.115pt}}
\multiput(1096.00,177.17)(19.807,-5.000){2}{\rule{1.010pt}{0.400pt}}
\multiput(1120.00,171.93)(2.602,-0.477){7}{\rule{2.020pt}{0.115pt}}
\multiput(1120.00,172.17)(19.807,-5.000){2}{\rule{1.010pt}{0.400pt}}
\multiput(1144.00,166.93)(2.602,-0.477){7}{\rule{2.020pt}{0.115pt}}
\multiput(1144.00,167.17)(19.807,-5.000){2}{\rule{1.010pt}{0.400pt}}
\multiput(1168.00,161.94)(3.552,-0.468){5}{\rule{2.600pt}{0.113pt}}
\multiput(1168.00,162.17)(19.604,-4.000){2}{\rule{1.300pt}{0.400pt}}
\multiput(1193.00,157.94)(3.406,-0.468){5}{\rule{2.500pt}{0.113pt}}
\multiput(1193.00,158.17)(18.811,-4.000){2}{\rule{1.250pt}{0.400pt}}
\multiput(1217.00,153.94)(3.406,-0.468){5}{\rule{2.500pt}{0.113pt}}
\multiput(1217.00,154.17)(18.811,-4.000){2}{\rule{1.250pt}{0.400pt}}
\multiput(1241.00,149.94)(3.552,-0.468){5}{\rule{2.600pt}{0.113pt}}
\multiput(1241.00,150.17)(19.604,-4.000){2}{\rule{1.300pt}{0.400pt}}
\multiput(1266.00,145.94)(3.406,-0.468){5}{\rule{2.500pt}{0.113pt}}
\multiput(1266.00,146.17)(18.811,-4.000){2}{\rule{1.250pt}{0.400pt}}
\multiput(1290.00,141.95)(5.151,-0.447){3}{\rule{3.300pt}{0.108pt}}
\multiput(1290.00,142.17)(17.151,-3.000){2}{\rule{1.650pt}{0.400pt}}
\multiput(1314.00,138.94)(3.552,-0.468){5}{\rule{2.600pt}{0.113pt}}
\multiput(1314.00,139.17)(19.604,-4.000){2}{\rule{1.300pt}{0.400pt}}
\multiput(1339.00,134.95)(5.151,-0.447){3}{\rule{3.300pt}{0.108pt}}
\multiput(1339.00,135.17)(17.151,-3.000){2}{\rule{1.650pt}{0.400pt}}
\put(1363,131.17){\rule{4.900pt}{0.400pt}}
\multiput(1363.00,132.17)(13.830,-2.000){2}{\rule{2.450pt}{0.400pt}}
\multiput(1387.00,129.95)(5.374,-0.447){3}{\rule{3.433pt}{0.108pt}}
\multiput(1387.00,130.17)(17.874,-3.000){2}{\rule{1.717pt}{0.400pt}}
\put(1412,126.17){\rule{4.900pt}{0.400pt}}
\multiput(1412.00,127.17)(13.830,-2.000){2}{\rule{2.450pt}{0.400pt}}
\end{picture}

Figure 4: $s_t, s_b, s_{\tau}$ against $\tan \beta$.
\end{center}
\noindent
Since we have assumed an $SU(5)$-type 
supersymmetric GUT with a gauge-Yukawa unification
in the third generation, the result (\ref{sum4}) 
is not a direct consequence
of a superstring model, although 
the form of the sum rules in both kinds of
unification schemes might coincide with 
each other as I mentioned (see footnote 11).
However, under the following circumstances (only rough), 
the sum rules (\ref{sum4})  could be a consequence of 
a superstring model:
(i) The Yukawa coupling of the third generation
is field-independent in the corresponding effective $N=1$
supergravity.
(ii) Below the string scale
an $SU(5)$-type gauge-Yukawa unification 
is realized so that
the sum rules  are
 RG invariant below the string scale
and are satisfied down to $M_{\rm GUT}$.
(iii) Below $M_{\rm GUT}$ the effective theory is the MSSM.

The sum rules (\ref{sum4}) 
could be experimentally
tested if the superpartners are found in future experiments,
 e.g., at LHC.
In any event, an experimental verification of the sum rules
of the SSB parameters would give an interesting information
on physics beyond the GUT scale.

\section{Conclusion}
Now I will come to conclusion.
Professor Zimmermann, obviously an interesting feature is coming.
So please keep staying in physics and
experience new developments in physics with us.

\vspace{1cm}
\noindent
{\bf Acknowledgment}

\vspace{0.3cm}
\noindent
I would like to thank the Theory Group
of the Max-Planck-Institute for Physics, Munich
for their warm hospitality which  my family and I have 
been enjoying since 1984.
I also would like to thank my collaborators,
Yoshiharu Kawamura, Tatsuo Kobayashi, Myriam Mondrag\' on,
Marek Olechowski, 
Klaus Sibold, Nicholas Tracas, George Zoupanos and Professor Wolfhart
Zimmermann. Most of the results I presented here were obtained 
in collaborations. I thank Professor Reinhard Oehme for
instructive  discussions.
At last but not least I would like to thank
Professor Wolfhart
Zimmermann for his  encouragement
of my research programs over the years.

\end{document}